\documentclass[nofootinbib,superscriptaddress,twocolumn]{revtex4-1}

\usepackage{amsmath}
\usepackage{amsfonts}
\usepackage{times}
\usepackage{graphicx}

\usepackage[utf8x]{inputenc}
\usepackage[T1]{fontenc}

\begin{document}


\bigskip

\title{Quantum clock: A critical discussion on space-time}

\author{Luciano Burderi}
\affiliation{\footnotesize{Dipartimento di Fisica, 
Universit\`a degli Studi di Cagliari, 
SP Monserrato-Sestu, KM 0.7, 09042 Monserrato, Italy}}
\author{Tiziana Di Salvo}
\affiliation{\footnotesize{Dipartimento di Fisica e Chimica, 
Universit\`a degli Studi di Palermo, 
via Archirafi 36, 90123 Palermo, Italy}}
\author{Rosario Iaria}
\affiliation{\footnotesize{Dipartimento di Fisica e Chimica, 
Universit\`a degli Studi di Palermo, 
via Archirafi 36, 90123 Palermo, Italy}}

\begin{abstract}
\noindent
We critically discuss the measure of very short time intervals.
By means of a {\it Gedankenexperiment}, we describe an ideal clock 
based on the occurrence of completely random events. 
Many previous thought 
experiments have suggested fundamental Planck-scale limits on 
measurements of distance and time.  Here we 
present a new type of thought experiment, based on a different type 
of clock, that provide further support for the existence of such limits.
We show that the minimum time interval $\Delta t$ that this clock 
can measure scales as the inverse of its size $\Delta r$. 
This implies an uncertainty relation between space and time:
$$
\Delta r \Delta t > G \hbar / c^{4},
$$ 
where $G$, $\hbar$, and $c$ are the gravitational constant, 
the reduced Planck constant, and the speed of light, respectively. 
We outline and briefly discuss the implications of this 
uncertainty conjecture.
\end{abstract}

\maketitle


\section{Introduction}
\label{intro}

The definition of a quantity or a concept in physics has to be 
operational in order to clarify the terms in which that quantity 
should be used and to avoid unjustified assumption of properties 
that belong more to our mental representation of that quantity 
than to its effective nature (e.g.~\cite{Bridgman27}).

This point of view has been particularly fruitful {\it e.g} when 
applied to the critical discussion of the concept of simultaneity, 
leading to the foundation of Special Relativity (SR, \cite{Einstein05}). 
Indeed, it is worth noting that an operational definition of 
time is crucial in SR. 
In particular the setting-up of a device that defines time 
in an operational way, whose behavior is constrained by the 
postulate of the invariance of the speed of light, 
implies directly the heterodox phenomenon of time dilation.
Such a device is the so called Light-Clock: two plane parallel
mirrors, facing each other at a constant along time -- {\it i.e.} fixed -- 
distance $\Delta x$, over which a light pulse bounces back and forth 
beating time in 
equal intervals of duration 
$\Delta t = \Delta x/c$, where $c$ is the speed of light.  

In what follows we adopt the rigorously operational 
definition of time as:
\begin{equation} 
\label{eq:timedef}
\begin{split}
&{\rm time \equiv a\; physical\; quantity\; that\; is\; measured\; 
by\; an }\\
&~~~~~~~~~~~~~{\rm appropriate\; clock}.
\end{split}
\end{equation} 
This apparently trivial (or somewhat circular) definition is essential 
to point out some subtle features of this elusive quantity.
The assumptions and the
limitations of any experimental apparatus adopted to measure (define) the quantity 
"time" have to be discussed carefully since they enter directly into play when the 
physical properties of the defined quantity enter into relationship with other 
physical quantities.

In particular, since in General Relativity (GR) time is a local quantity, deeply 
linked to every spatial point, it is desirable to keep the physical size of the
device used to measure it as small as possible, which results in the limitations 
discussed in \S \, \ref{relativity}.

To clearly address this question, in the next section we describe an ideal quantum 
device whose spatial extension can be suitably reduced, that is, in principle, capable 
of measuring arbitrarily short time intervals with any given accuracy. 
Curiously, this device is based on a process that, in some sense, is just the opposite 
of a strictly periodic phenomenon, namely the (in some respect more fundamental) 
occurrence of totally random events, such as the decay of an ensemble of 
non-interacting particles in an excited state. 
In this case the time elapsed may be obtained by the amount 
of particles that have decayed. Such a device has been discussed 
in~\cite{Salecker58} as an example of a simple microscopic clock.
We dubbed this device ``Quantum Clock''.
Limits, imposed on our device by Quantum Mechanics (QM) and GR, 
result in an uncertainty relation that we briefly discuss.

Many previous thought experiments have suggested fundamental Planck-scale
limits on measurements of distance and time (see {\it e.g.}~\cite{Hossenfelder12} for a review). 
Here, we present a new set of thought experiments, based on a different type of clock (which
was briefly alluded to in~\cite{Salecker58}), that provide further
support for the existence of such limits.

\section{The Quantum Clock}
\label{quantumclock1}

Let us consider a statistical process whose probability of occurrence
\begin{equation} 
\label{eq:random}
dP = \lambda \; dt
\end{equation} 
is independent of time ({\it i.e.} $\lambda$ constant with time).

A good example of this sort of situation is given by radioactive decay
\footnote{
The exponential decay law implied by equation (\ref{eq:random}) is, 
strictly speaking false for an unstable quantum system
(see {\it e.g.}~\cite{Dicus02} and references therein). 
In particular the deviations at very short times are known as quantum Zeno effect. 
On the other hand, Fonda, Ghirardi \& Rimini \cite{Fonda78}, 
with a formalism based on a definition of an unstable state that takes into 
account ``the fact that an unstable system unavoidably interacts with its 
environment'' (to use their own words), demonstrated that, from the experimental 
point of view (in the sense of its definition given by Operationalism),
the resulting non--decay probability is, for all practical purposes, a pure exponential, 
although there is a possible dependence of the decay lifetime on the experimental 
apparatus.
These modifications of the simple exponential decay law result, in general,  into an 
increased effective lifetime (or shortened decay rate) with respect to the theoretical 
value (calculated from the moment in which the unstable system 
is prepared), as compared to the theoretical lifetime,
which, ultimately, 
strengthens the inequality that we propose in this paper: equation 
(\ref{eq:spacetime}) of \S \, \ref{relativity}.
}.
Given an amount of radioactive matter of mass $M = N m_{\mathrm p}$, 
where $N$ is the number of particles and $m_{\mathrm p}$ is the mass 
of a single particle, it can be 
easily proved that, if equation (\ref{eq:random}) holds, 
the mean variation of the number of particles in the unit time interval 
is given by
\begin{equation}
\label{eq:decay}
\frac{dN}{dt} = - \lambda N
\end{equation}
(see {\it e.g.}~\cite{Bevington69}).
The mean number of decays in a time $\Delta t \ll \lambda^{-1}$ is 
$\Delta N_{\Delta t} = \lambda N \Delta t$. 
The measured number of decays fluctuates around the expected value 
with Poissonian statistics 
{\it i.e} with $\sigma = \sqrt{\lambda N \Delta t}$. 
Therefore it is possible to measure a time interval $\Delta t$ counting 
$\lambda N \Delta t$ 
events. The relative error on our measure is 
$\sigma_t / \Delta t = \epsilon = (\lambda N \Delta t)^{-1/2}$.
Whenever it is required to measure a short time interval
($\Delta t \rightarrow 0$), 
in order to keep the relative error below a given threshold, say 
$\sigma_t / \Delta t < \epsilon_0$, 
one should keep the product $\lambda N \Delta t > \epsilon_0^{-2}$. 
Providing that enough particles are available, 
$N$ can be conveniently increased up to the required precision. 

A physical device based on the process discussed above can be built 
in several ways. 
The simplest (albeit perhaps not practical) device consists
of a given amount of mass $M$ of radioactive particles 
(corresponding to a given 
number of particles $N = M/m_{\mathrm p}$) ,
that decay by emitting a photon,
completely surrounded by 
proportional counters ({\it e.g.} Geiger--Muller counters) of quantum 
efficiency $\sim 1$. We consider this device, dubbed 
Quantum Clock hereinafter (QmCl), for 
the operational definition of time. If we count a number of decays 
${\cal N}_{\Delta t} \sim \Delta N_{\Delta t} = \lambda N \Delta t$ 
in the QmCl the time elapsed is
\begin{equation}
\label{eq:1clock}
\Delta t  = \frac{{\cal N}_{\Delta t}m_{\mathrm p} }{\lambda M}
\end{equation}
where we have expressed the number of particles in terms of their mass.
The associated relative accuracy is $\epsilon = \sigma_t / 
\Delta t = {\cal N}_{\Delta t}^{-1/2}$. Because at least one event 
must be recorded by the device we have $ \epsilon \le 1$. 
Therefore in terms of this uncertainty, and
expressing $m_{\mathrm p}$ in terms of its rest energy $E_{\mathrm p} =
m_{\mathrm p} c^{2}$, the time elapsed is
\begin{equation}
\label{eq:2clock}
\Delta t  = \frac{1}{\epsilon^{2} M c^{2}} \times 
\frac{E_{\mathrm p}}{\lambda}.
\end{equation} 

\section{The Quantum Clock and the Heisenberg Uncertainty Principle}
\label{heisenberg}

As a quantum device, the QmCl is subject to Heisenberg's uncertainty relations.
In particular we will use the relation between energy and 
time, namely 
\begin{equation}
\label{eq:heis}
\delta E \times \delta t \ge \hbar / 2
\end{equation}
where $\delta E$ and $\delta t$ are the uncertainties in 
energy and time coordinate of a particle, and $\hbar = h/(2\pi)$ is the reduced Planck constant.

In (\ref{eq:2clock}) the factor $E_{\mathrm p} / \lambda$ depends 
on the specific
nature of the radioactive substance used for the construction of 
the clock.
To make the QmCl independent of the particular substance adopted, 
we consider the limitations imposed by (\ref{eq:heis}).
Let us consider a particle of energy $E_{\mathrm p}$.  
Conservation of mass-energy imposes an upper limit to the
uncertainty in the decay energy $\delta E_{\mathrm{p}}$ namely
$E_{\mathrm{p}} \ge \delta E_{\mathrm{p \; max}}$
where $\delta E_{\mathrm{p \; max}}$ is the maximum uncertainty obtainable
in the measure of the decay energy.
The decay rate $\lambda$ is the inverse of the average decay time
$1/\lambda = \tau  \ge \delta t_{\mathrm{min}}$ 
where $\delta t_{\mathrm{min}}$ is the minimum uncertainty obtainable
in the measure of the time elapsed before the decay.
Since the maximum uncertainty in the energy and the minimum 
uncertainty in the time elapsed are related by the uncertainty 
relation (\ref{eq:heis}), we have:
\begin{equation}
\label{eq:2heis}
\frac{E_{\mathrm{p}}}{\lambda} = E_{\mathrm{p}} \times \tau \ge
\delta E_{\mathrm{p \; max}} \times \delta t_{\mathrm{min}} \ge \hbar / 2
\end{equation}
Inserting (\ref{eq:2heis}) in (\ref{eq:2clock}) we have
\begin{equation}
\label{eq:3clock}
\Delta t  \ge \frac{\hbar}{2 \epsilon^{2} c^{2}} \times \frac{1}{M}
\end{equation}
which expresses the same lower limit for the mass of a clock, 
capable to measure time intervals down to an accuracy $\Delta t$, 
given by~\cite{Salecker58} (see eq. 6 in their paper).
More recently, Ng and van Dam (\cite{Ng03}, and reference therein) 
discussed a similar relation which limits the precision
of an ideal clock (see eq. 8 in their paper).
In the above relation the ``fuzziness'' of the QM manifest itself in the
inequality. However the mass in the denominator of the second member allows,
in principle, to build such a massive clock that an arbitrarily short time
interval can be adequately measured with the required accuracy $\epsilon$.

\section{The Quantum Clock and General Relativity}
\label{relativity}

In GR time is a {\it local} quantity in the sense that, 
in general, the metric implies time coordinate factors 
(namely the factors in front of the time coordinate interval)
which may be different
in different points of space.
If a non-uniform gravitational field is present
({\it e.g.} the gravitational field outside a spherical mass distribution, Schwarzschild metric),  
the proper time of an observer at rest in a place where the gravitational field is quite intense 
({\it e.g.} at a few  Schwarzschild's radii from the centre of the mass distribution),
is shorter than the time interval measured by a clock located where the gravitational field is less intense 
({\it i.e.} at several Schwarzschild's radii from the 
centre of the mass distribution).
Because of its spatial extension, a clock defined by (\ref{eq:3clock}) 
is capable of measuring a sort of ``average'' time interval over the region 
defined by its size. Since we measure time counting events which may
occur randomly in any point of the clock, the size of the region over 
which we are measuring the time is identified with the entire size
of the clock. In other words a spatial uncertainty, corresponding to 
the finite size of the clock, is associated with the measure 
of time. To minimize this uncertainty (providing that the single particles
are so weakly interacting with each other that their behavior is unaffected 
by the proximity of neighbors) it is possible to compress the QmCl 
in order to make its spatial extension as small as possible. 

However the presence of the clock mass affects the structure of space-time. 
In particular, the {\it Hoop Conjecture} (see e.g.~\cite{Misner73})
states that a piece of matter of mass $M$, around which -- in every direction -- it is possible to place
a circular hoop of length $2 \pi R_{\rm Sch}$, unavoidably undergoes gravitational 
collapse ($R_{\rm Sch} = 2GM/c^2$ is the Schwarzschild radius, $G$ is the gravitational constant).

In the following, we will assume that the volume in which 
the QmCl is compressed is spherically symmetric and coincident with a sphere.
The assumption of a spherical clock is not merely a simplification, but an 
important issue whose consequences will be discussed 
in \S \, \ref{discussion}. 

If the QmCl were compressed in a spherical volume whose circumferential 
radius is smaller than its Schwarzschild radius
it would face the gravitational collapse and it would not be useable 
as a device to measure time intervals because the products of the decays
({\it e.g.} the photons of the example discussed above) 
cannot escape outside the Schwarzschild radius to bring the information 
that time is flowing in that region of space.

This implies that the smallest possible radius for the QmCl is its 
Schwarzschild radius, $R > R_{\rm Sch}$ or 
\begin{equation}
\label{eq:radius}
\frac{1}{M} > \frac{2 G}{c^2 R}.
\end{equation}
The condition above has been discussed in the literature as a necessary
lower limit on the size of a massive clock. In particular Amelino-Camelia
proposed an equation for a lower bound on the uncertainty in the measurement
of a distance in which the condition above is included~\cite{Camelia94}, 
\cite{Camelia96}.

Inserting (\ref{eq:radius}) into (\ref{eq:3clock}) gives
\begin{equation}
\label{eq:1uncertainty}
\Delta t R > \frac{1}{\epsilon^2}
\frac{G \hbar}{c^{4}},
\end{equation} 
where $R = \Delta r$ is the radius of the QmCl 
({\it i.e.} the uncertainty on the exact position of the radioactive
decay events).
Because, as we noted in \S \, \ref{quantumclock1}, $\epsilon \le 1$, 
we can write 
\begin{equation}
\label{eq:spacetime}
 \Delta r \Delta t > \frac{G \hbar}{c^{4}} 
\end{equation}
The equation above quantifies the impossibility to simultaneously 
determine spatial and temporal coordinates of an event with
arbitrary accuracy.
\section{Discussion}
\label{discussion}

Several thought experiments have been proposed to explore fundamental 
limits in the measurement process of time and space intervals 
(see {\it e.g.}~\cite{Hossenfelder12} for an updated and 
complete review). In particular Mead~\cite{Mead64} ``postulate the existence
of a fundamental length'' (to use his own words) and discussed the possibility
that this length is the Planck length, $\ell_{\rm min} \sim \sqrt{G \hbar/c^3} = \ell_{\rm Planck}$, which resulted in 
limitations in the measure of arbitrarily short time intervals originating 
relations similar to (\ref{eq:spacetime}) of \S \, \ref{relativity}. 
Moreover in a subsequent paper  \cite{Mead64}, Mead discussed an in principle 
observable spectral broadening, consequence of the postulate the existence 
of a fundamental length of the order of Planck Length.
More recently, in the framework of String Theory a space-time 
uncertainty relation
has been proposed which has the same structure of the uncertainty relation 
discussed in this paper (\cite{Yoneya87},~\cite{Yoneya89}, see {\it e.g.}
\cite{Yoneya97} for a discussion of
the possible role of a space-time uncertainty relation in String Theory). 
The relation proposed in String Theory 
constraints the product of the uncertainties in the time interval $c \Delta T$
and the spatial length $\Delta X_l$ to be larger than the square of the 
string length $\ell_S$, which is a parameter of the String Theory. 
However, to use the same words of Yoneya \cite{Yoneya97}, this relation is 
``speculative and hence rather vague yet''. 
Indeed, in the context of Field Theories, uncertainty relations 
between space and time coordinates similar to that proposed here
have been discussed as an ansatz for the limitation arising in combining 
Heisenberg's uncertainty principle with Einstein's theory of 
gravity~\cite{Doplicher95}.
In 1995 Garay~\cite{Garay95} postulated and discussed, in the context of Quantum Gravity,
the existence of a minimum length of the order of the Planck Length, but followed the idea
that this limitation may have a similar meaning to the speed limit defined by the speed
of light in Special Relativity, in line with what was already pointed out previously
(see {\it e.g.}~\cite{Borzes88} and references therein). 
In the framework of the so called Quantum Loop Gravity (see {\it e.g.}~\cite{Rovelli88a},~\cite{Rovelli88b} 
and~\cite{Rovelli98} for a review) a minimal length appears characteristically in the form
of a minimal surface area (\cite{Rovelli95},~\cite{Astekhar97}): indeed the area
operator is quantized in units of $\ell_{\rm Planck}^2$~\cite{Rovelli93}.
It has been sometimes argued that this minimal length might conflict with Lorentz invariance, 
because a boosted observer could see the minimal length further Lorentz contracted, 
however this problem is solved in Loop Quantum Gravity since the minimal area 
does not appear as a fixed property of geometry, but rather as the minimal (nonzero) eigenvalue of a 
quantum observable that has the same minimal area $\ell_{\rm Planck}^2$ for all the
boosted observers. What changes continuously in the boost transformation is the probability distribution 
of seeing one or the other of the discrete eigenvalues of the area (see {\it e.g.}~\cite{Rovelli02}).

In this paper the analysis of a thought experiment designed to explore fundamental 
limits of clocks has brought us to consider the uncertainty
relation between space and time expressed in equation (\ref{eq:spacetime}).
Here we briefly outline some of the implications of this equation. 
To this aim it is useful to represent space (indeed circumferential distance since the "size" of the
QmCl is limited by assuming the {\it Hoop Conjecture}) and time intervals 
in a standard space-time diagram.
We choose the space and time units in order to have $c =1$, or 
$c \Delta t$ as the ordinate. In this representation the
bisector defines the null intervals, separating the timelike intervals,
above the bisector, from the spacelike intervals, below.
The relation (\ref{eq:spacetime}) of \S \ref{relativity}, namely
\begin{equation}
\label{eq:2spacetime}
\Delta r \, c \Delta t > \frac{G \hbar}{c^{3}} 
\end{equation}
defines an hyperbola in this plane whose asymptotes are the $\Delta r$
and $c \Delta t$ axes and whose vertex is located at 
$(\Delta r)_{\rm vertex} = (c \Delta t)_{\rm vertex} = \sqrt{G \hbar/c^3}$. 

The following considerations can be made: \\
i) The minimum (measurable) spatial circumferential distance is the Planck Length. 
This is because a proper space distance is defined 
for spacelike intervals and the minimum circumferential distance coordinate ``$\Delta r$'' of the points 
below the bisector is $(\Delta r)_{\rm min} = \sqrt{G \hbar/c^3}$, which
is the Planck Length.
It is important to note that the assumption of a QmCl of spherical shape,
made \S \, \ref{relativity}, means that, in principle, it is possible to 
measure a spatial coordinate with an accuracy better than the Planck Length.
In particular, it is possible to measure the coordinate position of an 
object ({\it e.g.} the $x$ coordinate of its centre of mass) to much better 
precision if the object is big enough and extended enough along the other 
coordinates ({\it i.e.} in the $y$--$z$ plane). 
Indeed it is conceivable to build an object (our clock) with different length 
in each dimension, say $\Delta x$, $\Delta y$, and $\Delta z$, 
and to violate the uncertainty relation proposed for one of these lengths. 
However a firm upper limit should exist since the {\it Hoop Conjecture} 
establishes that the collapse of the object/clock should be 
unavoidable once $\Delta r = \sqrt{\Delta x^2 +\Delta y^2 +
\Delta z^2 } \leq 2GM/c^2$, where $M$ is the mass of the clock/object.
Thus, the assumption of a QmCl of spherical shape, shows that is impossible
to measure a ``size'' $\Delta r \equiv \sqrt{\Delta x^2 +\Delta y^2 +
\Delta z^2 }$ smaller than Planck length
\footnote{We thank L. Susskind (private communication) for pointing out this subtle
although important question.}. \\
ii) The minimum (measurable) time interval is the Planck Time. 
This is because a proper time interval is defined 
for timelike intervals and the minimum ``$\Delta t$'' coordinate of the points 
above the bisector is $(\Delta t)_{\rm min} = \sqrt{G \hbar/c^5}$, which is
the Planck Time. \\
iii) The uncertainty relation 
cannot be violated by using the phenomenon of length contraction 
predicted by Special Relativity.
As the spatial length in the direction of motion (with speed 
modulus $v$)   
is contracted by the inverse of the Lorentz factor 
$\gamma^{-1} = [1 - (v/c)^2]^{1/2}$, it would in
principle be possible 
to imagine to build a clock capable of measuring time intervals
of duration $\Delta t$ whose proper length along a 
given direction, say the $x$ axis, is such that its size
($\sqrt{\Delta x^2 +\Delta y^2 +
\Delta z^2 }$ ) is slightly above the minimum $\Delta r$ determined by (\ref{eq:spacetime}).
It is then possible to observe the clock from a reference system in which the clock 
moves at uniform speed $v$ along the $x$ axis. 
If the speed is high enough, the length of the clock in the direction of motion
is so Lorentz contracted that its size falls below the minimum above. 
Time dilation, by the Lorentz factor $\gamma = [1 - (v/c)^2]^{-1/2}$, 
prevents 
the violation of the uncertainty relation. \\
iv) The uncertainty relation holds in Schwarzschild metric~\cite{Schwarz16} since: 
\begin{eqnarray}
\label{eq:schwarzscild}
ds^2 = (1 - R_{\rm Sch}/r) c^2 dt^2 - (1 - R_{\rm Sch}/r)^{-1} dr^2 
\nonumber \\
- r^2 (d\theta^2 + \sin^2\theta d\phi^2),
\end{eqnarray}
 where $ds$ is the infinitesimal interval, 
 $r$ is the circumferential radius, $\theta$ and $\phi$
 are the angles of a spherical coordinate system.

Although relation (\ref{eq:spacetime}) has the structure of an uncertainty
relation, and therefore does not contain a minimum spatial length or a 
minimum time duration explicitly, the timelike and spacelike classification
of the intervals, determined by Special Relativity, when combined with 
(\ref{eq:spacetime}), implies (in a somewhat unexpected way) the 
existence of minimal space-time ``quanta'' equal to the product of the Planck 
Length and Time, respectively. 
In other words, equation (\ref{eq:spacetime}) would mean that
Nature manifest the existence of ``atoms'' of space and time (whose size 
does not require the introduction of any extra parameter in the theory)  
only whenever space and time are simultaneously probed down to the 
smallest scale. No intrinsic discreteness characterizes, individually, 
space or time coordinates, which are the continuous and smooth components of 
the fabric of space-time. Discreteness naturally emerges 
in the operational definition of a simultaneous measure of 
space and time whether the limitations imposed by the Uncertainty Principles of QM and the 
formation of an Event Horizon (occurring -- according to GR -- during complete gravitational 
collapse) are taken into account, together with the universal hyperbolic character of the 
``local'' metric implied by SR.
Indeed QM, GR, and SR enter in the uncertainty relation through their 
fundamental constants, $\hbar$, $G$, and $c$. 

We finally note that in the limits $\hbar \rightarrow 0$ and 
$c \rightarrow \infty$ (that means in the classical limit), there is no 
uncertainty relation between space and time, as expected. A similar 
discussion on the role of $\hbar$, $c$, and $G$ in determining a spatial 
resolution limit can be found in~\cite{Garay95}.

\section{Conclusions}
\label{conclusions}

In this paper, by means of a {\it gedankenexperiment}, 
we have argued on the existence of a Space-Time uncertainty relation
expressed by equation (\ref{eq:spacetime}).
This relation does not depend 
on parameters defined within a specific theory but only on fundamental 
constants, since the previously  discussed "free" parameter of String Theory $\ell_S^2$ is 
naturally replaced by $G \hbar/c^3$ in our QmCl, 
without postulating the existence of any fundamental length. 
We argue that an uncertainty relation between space and time 
arises naturally if we take into account: i) the Uncertainty Principles 
of Quantum Mechanics; ii) the well known result of GR that gravitational 
collapse -- and the subsequent formation of an Event Horizon -- 
is unavoidable once a given amount of mass-energy is concentrated 
into a spatial extension smaller than the volume encompassed by the 
Schwarzschild Radius of that amount of mass-energy.
Fundamental (minimal) space length and time interval emerge in a natural way, 
when this uncertainty relation is considered within the local Minkowskian
structure of space-time.

Canonical commutation relations imply the Heisenberg Uncertainty 
Principles~\cite{Kennard27}: the product of the standard deviations of two 
operators that do not commute is greater or equal to the modulus
of the average value of the operator built from the commutator
of the two operators~\cite{Robertson29}. 
It is therefore possible to develop QM  adopting the Heisenberg uncertainty 
principles (and the commutation relations associated) as the postulates over 
which the whole quantum theory is built.
This justifies the view, shared by many physicists, that the Heisenberg uncertainty 
relations should be considered as fundamental laws of Nature (see {\it e.g.}
\cite{Baggott92}).
In a similar way, we argue that it would be possible to develop 
the foundations of a mathematical theory of gravity, which will be fully 
consistent with the postulates of the quantum theory, starting from the
uncertainty relation between space and time discussed above. 

\begin{acknowledgments}
First, we would like to thank Giovanni Amelino-Camelia and Leonard Susskind for several 
stimulating discussions on the topics discussed in this paper which have deepened our 
understanding of the subject. We are also grateful to the referees Steven Carlip and Thibault Damour 
for useful discussions and suggestions that have greatly improved quality of the present paper. Finally, 
we would like to thank many colleagues and friends who have contributed with encouraging and stimulating 
discussions that helped us in clarifying several subtle points addressed in this work: among the others, 
F. Burderi, L. Colombo, E. Fiordilino, F. Haardt, N. La Barbera, M.T. Menna, R. Passante, L. Rizzuto, C. Rovelli, 
L. Stella. 
This work was partially supported by the Regione Autonoma della Sardegna through POR-FSE Sardegna 2007-2013, L.R. 7/2007, 
Progetti di Ricerca di Base e Orientata, Project No. CRP-60529, and by the Fondo Finalizzato alla Ricerca (FFR) 2012/13, 
Project No. 2012-ATE-0390, founded by the University of Palermo.
\end{acknowledgments}

\end{document}